\documentclass[prc,aps,nofootinbib,showkeys,showpacs,twocolumn]{revtex4}   
\usepackage{epsfig}  
\usepackage{graphicx}  
\usepackage{amssymb}
\usepackage{color}  
\begin{document}  
  
\title{Systematic of isovector and isoscalar giant quadrupole resonances in normal and superfluid 
spherical nuclei}

\author{Guillaume Scamps}  
 \email{scamps@ganil.fr}  
\affiliation{GANIL, CEA/DSM and CNRS/IN2P3, Bo\^ite Postale 55027, 14076 Caen Cedex, France}  
\author{Denis Lacroix} \email{lacroix@ganil.fr}  
\affiliation{Institut de Physique Nucl\'eaire, IN2P3-CNRS, Universit\'e Paris-Sud, F-91406 Orsay Cedex, France}  
\affiliation{GANIL, CEA/DSM and CNRS/IN2P3, Bo\^ite Postale 55027, 14076 Caen Cedex, France}  
    
\begin{abstract}  
The isoscalar (IS) and isovector (IV) quadrupole responses of nuclei are systematically investigated using the time-dependent 
Skyrme Energy Density Functional including pairing in the BCS approximation. 
Using two different Skyrme functionals, Sly4 and SkM*, respectively 263 and 304 nuclei have been found to be 
spherical along the nuclear charts. The time-dependent evolution of these nuclei has been systematically 
performed giving access to their quadrupole responses. It is shown that the mean-energy of the collective high energy 
state globally reproduces the experimental IS and IV collective energy but fails to reproduce their lifetimes.  
It is found that the mean collective energy depends rather significantly 
on the functional used in the mean-field channel. Pairing by competing with parity effects can slightly affect 
the collective response around magic numbers and induces a reduction of the collective energy compared to the average trend. 
Low-lying states, that can only be considered if pairing 
is included, are investigated. While the approach provides a fair estimate of the low lying state energy, it strongly underestimates the transition rate $B(E2)$. Finally, the possibility to access to the density dependence of 
the symmetry energy through parallel measurements of both the IS- and IV-GQR is discussed.
\end{abstract}

\keywords{collective motion, TDHF, pairing}
\pacs{21.10.Pc, 21.60.Jz , 27.60.+j}
  
\maketitle  
  
\section{Introduction}  

In recent years, large efforts have been devoted within the energy 
density functional (EDF) to bring time-dependent theories at the level of state of the 
art nuclear structure theories.  A current topic of interest is to include pairing correlations
in nuclear dynamics \cite{Has07,Ave08,Ste11,Eba10,Sca12,Sca13}. In the present work, we systematically 
investigated the possibility to describe the isoscalar (IS) and isovector (IV)-GQR including 
pairing using  the Time Dependent Hartree-Fock +BCS (TDHF+BCS) approach. There are several reasons to have chosen the GQR has 
a test ground (i) as far as we know, the isoscalar and isovector GQR have never been systematically investigated using a 
time-dependent EDF framework and therefore our study should make possible to identify weakness and strength of 
this approach in this context; (ii) a systematic analysis exists using the QRPA approach allowing for comparisons 
\cite{Ter06}; (iii) a large number of experimental data exists especially for the IS \cite{Ber81,Li10,Lui11,Bue84,You81,Bor89,Sha88,Bra85} and therefore not only 
qualitative but also quantitative studies can be made (iv) the experimental observation 
of the IV-GQR has recently made some progress \cite{Sim97,Ich02,Hen11}; (v) thanks to these progress, the GQR 
has been recently proposed as a possible alternative tool to get informations on the density dependence of the symmetry energy \cite{Roc13} 

In the present work, we have systematically investigated the evolution of nuclei that are found 
spherical in their ground states using two different Skyrme functionals. The static properties and the dynamical evolution have been obtained consistently 
using the HF+BCS and TDHF+BCS approach. The TDHF+BCS method, while less general than the 
TDHFB framework and leading to specific difficulties \cite{Sca12}, has the advantage that each evolution can be made in a reasonable time allowing 
to perform the calculation over a wide range of nuclei.  

In the following, the microscopic approach is briefly described. Then, the   protocol for selecting nuclei 
is discussed as well as some average properties related to the pairing correlations, the root mean-square radius...
Properties of the response in the low-lying and high-lying sector are investigated.  We finally illustrate the usefulness of large scale GQR study with the possibility to infer from it information on the symmetry 
energy. 

\section{The TDHF+BCS approach to giant resonances}

The TDHF+BCS theory is a simplified version of the TDHFB approach where the off-diagonal 
part of the pairing field is neglected. Properties, advantages and drawback 
of this theory have been discussed extensively in Refs. \cite{Eba10,Sca12,Sca13} 
and we only give below a minimal summary of important aspects to treat giant resonances.

For a given nucleus, the initial wave-function is obtained using the {\rm EV8} code \cite{Bon05} 
that solves the HF+BCS in r-space with the Skyrme functional in the mean-field term and with a contact 
interaction in the pairing channel. After, this step, the N-body wave-packets takes the form of a quasi-particle 
vacuum written in a BCS form as:
 \begin{eqnarray}
| \Psi_0 \rangle &=& \prod_{k>0} \left( u^0_k + v^0_k a^\dagger_k a^\dagger_{\bar  k} \right)|-\rangle.  \label{eq:fullwf}
\end{eqnarray} 
where $u^0_k$ and $v^0_k$ are usual components of the quasi-particle states while $a^\dagger_k$ and 
$a^\dagger_{\bar  k}$ are the creation operators of time-reversed single-particle states, denoted respectively by 
$| \varphi_k \rangle$ and $| \varphi_{\bar k} \rangle$.   

It could be shown \cite{Eba10}, that the state  (\ref{eq:fullwf}) is a stationary solution of the TDHF+BCS set of coupled equations
given by:
\begin{eqnarray}
\left\{
\begin{array} {l}  
i\hbar \partial_t | \varphi_k \rangle = (h[\rho](t)-\eta_k(t)) | \varphi_k \rangle  , \\
\\
i \hbar \dot n_k(t) =  
\kappa_{k}(t) \Delta_{k}^*(t) - \kappa_{k}^*(t) \Delta_{k} (t), \\  
\\
i \hbar \dot \kappa_k(t) =   
\kappa_k(t) ( \eta_k(t) + \eta_{\overline{k}}(t) ) + \Delta_{k}(t) (2n_k(t)-1) ,
\end{array}
\right. \label{eq:tdhfbcs}
\end{eqnarray}  
where $\eta_k(t) = \langle \varphi_k(t) |h[\rho]| \varphi_k(t) \rangle$  
and where $n_k(t)= |v_k(t)|^2$  and $\kappa_k(t)=u_k^* (t) v_k (t) $
are respectively the occupation numbers and anomalous density components. 
 
To study giant resonances, we follow the method standardly used in time-dependent 
approaches (see for instance \cite{Sim12}) and apply a small initial boost to the 
wave-packet $| \Psi_0 \rangle$:
 \begin{eqnarray}
| \Psi (t=0) \rangle &=& e^{-i \lambda \hat Q}  | \Psi_0 \rangle, \label{eq:boost}
\end{eqnarray} 
where $\lambda$ is a number that is small enough to insure the validity of the linear approximation. 
$\hat Q$ here is an operator that depends on the type of modes one would like to study. in the present work, 
we are interested in the giant quadrupole response in spherical nuclei where the excitation operators are given by \cite{Ter06}
\begin{eqnarray}
\hat Q^{IS}_{20} &=& e \sum_{i=1}^{A} {r^2_i} Y_{20}(\Omega_i), \label{eq:qis} \\
\hat Q^{IV}_{20} &=& \frac{eN}{A} \sum_{i=1}^{Z} {r^2_i} Y_{20}(\Omega_i) - \frac{eZ}{A} \sum_{i=1}^{N} {r^2_i} Y_{20}(\Omega_i), \label{eq:qiv}
\end{eqnarray}
respectively in the isoscalar and isovector channel and where $A$, $N$ and $Z$ are respectively the total, neutron and proton 
numbers.

The boost (\ref{eq:boost}) induces a local boost to the single-particle components while 
leaving  the initial $(u_k,v_k)$ components unchanged, i.e. $u_k(t_0) = u^0_k$ and $v_k(t_0) = v^0_k$. 
The response of the nucleus to the initial perturbation is studied by solving the coupled set of 
TDHF+BCS equations in time. The strength function can then be obtained by performing the Fourier transform 
of the evolution of the operator $\hat Q$. More precisely, we have \cite{Eba10,Sim12}
\begin{eqnarray}
S_Q(\omega) &=& \lim_{\lambda \rightarrow 0} -\frac{1}{\pi \lambda \hbar } \int^{\infty}_{t_0} dt \left( Q (t)- Q(t_0) \right) \sin(\omega t), 
\label{eq:strength}
\end{eqnarray}   
where 
\begin{eqnarray}
Q(t) \equiv \langle \Psi(t) |\hat Q| \Psi(t) \rangle = \sum_{k} \langle \varphi_k(t) |\hat Q | \varphi_k(t)\rangle n_k(t).
\end{eqnarray} 
Note that here, only spherical nuclei are considered and $Q(t_0) =0$. 

In practice, the time evolution cannot be performed up to infinite time and  a damping factor is assumed, i.e. $ \sin(\omega t) \rightarrow e^{-\Gamma_0 t/ 2\hbar} \sin(\omega t)$ in formula (\ref{eq:strength}). Then, decomposing the small amplitude vibration on the eigenstates of the system, denoted 
by $| \Psi_\nu \rangle$ associated to energy $\hbar \omega_\nu = E_\nu - E_0$, $E_0$ being the ground state energy, the strength function writes: 
\begin{eqnarray}
S_Q(E) &=& \frac{1}{2\pi} \sum_\nu |\langle \Psi_0| \hat Q | \Psi_\nu \rangle |^2 \nonumber \\
&& \times \left\{ \frac{\Gamma_0} {(E-\hbar \omega_\nu)^2 + \Gamma_0^2/4 } - \frac{\Gamma_0} {(E+\hbar \omega_\nu)^2 + \Gamma_0^2/4 } \right\}.
\label{eq:strengthG}
\end{eqnarray}   
In the zero damping limit, one recover the standard form of the strength function generally used in RPA or QRPA \cite{Rin80}. Note that in the following, we will simply use the notation $S_{IS}$ and $S_{IV}$ for the isoscalar and isovector strength function respectively. 

 \subsection{Selection of Nuclei}
 \begin{figure}[!ht]   
	\centering\includegraphics[width=\linewidth]{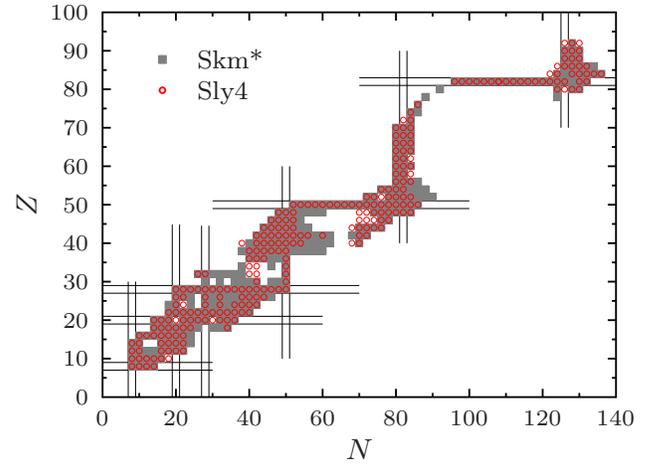}  
	\caption{(Color online) Spherical nuclei retained in the present study using Sly4 (red open circles) and SkM* 
	(grey filled squares). The 
	horizontal and verticals lines indicates protons and neutrons magic numbers. } 
	\label{fig:map_spher} 
\end{figure} 
In the present study, we have systematically considered initial nuclei that are found to be spherical
in the {\rm EV8} code. The mesh size has been taken as $\Delta x =\Delta y = \Delta z  = 0.8$ fm and the total 
size of the mesh are $2L_x = 2L_y = 2L_z = 22.4$ fm. Two different functionals, namely the 
Sly4 \cite{Cha98} and the SkM* \cite{Bar82} have been employed in the mean-field channel. 
These two functionals have the advantage to be widely used in nuclei allowing for comparison 
of the present results with other approaches like the QRPA of Ref. \cite{Ter06}. Only 
proton-proton and neutron-neutron pairing is considered. The pairing effective interaction is given by 
\begin{eqnarray}
V_\tau ({\mathbf r}, \sigma; {\mathbf r'} ,\sigma') &=& V^{\tau \tau}_0 \left(1 - \eta \frac{\rho([{\mathbf r}+{\mathbf r'}]/2)}{\rho_0} \right)\delta_{{\mathbf r},{\mathbf r'}}
\left[1-P_{\sigma
\sigma'} \right], \nonumber
\end{eqnarray}   
where $P_{\sigma \sigma'}$ is the spin exchange operator and where $\rho_0 = 0.16$ fm$^{-3}$ and $\tau=n,p$.
Results presented below will be presented only for the case of surface interaction 
with parameters given in table I of ref. \cite{Ber06}.
The selection of nuclei is done starting from a list of 749 even-even nuclei for which the masses have been 
measured and which have a mass $A \ge 8$. 
In the {\rm EV8} code, the single-particle states of the Nilsson hamiltonian are first obtained to initiate the 
imaginary time convergence. For each nuclei, to avoid the convergence towards local minima that sometimes 
might happen with {\rm EV8}, 
we performed 4 types of Nilsson calculations to initiate the self-consistent convergence: one assuming spherical shape, one oblate, one prolate and one triaxial deformation. 
Nuclei are assumed to be spherical if the deformation parameter after the imaginary time verifies $\beta_2<0.001$ where 
the deformation parameter is defined through 
\begin{eqnarray}
 \beta_2 &=& \sqrt{\frac{5}{16 \pi}} ~ \frac{4 \pi}{3R^2 A} ~ \langle \hat Q^{IS}_{20} \rangle \, , \label{eq:defor}
\end{eqnarray}
with $R=1.2~ A^{1/3}$. Altogether,  
304 and 263 nuclei have been found to be spherical with SkM* and with Sly4 respectively. 
The retained nuclei are shown  in Fig. \ref{fig:map_spher}.
 \begin{figure}[!ht] 
	\centering\includegraphics[width=\linewidth]{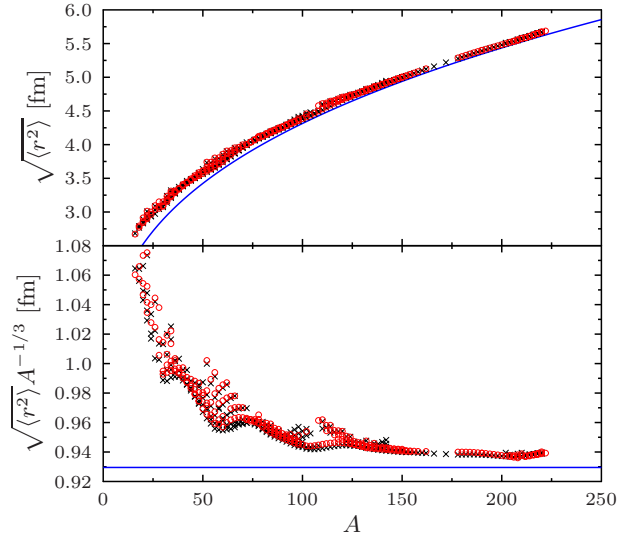}  
	\caption{(Color online) Top: Root-mean square radius $\sqrt{\langle r^2 \rangle}$ obtained for the selected spherical nuclei using the 
	 Sly4 (red circles) or SkM* (black crosses) functional. The simple approximation $\sqrt{\langle r^2 \rangle} = 
	 \sqrt{\frac35} (1.2 A^{1/3})$ is also shown as a reference with solid blue line.  Bottom: Same quantity divided 
	 by $A^{1/3}$. } 
	\label{fig:comp_rms} 
\end{figure} 
\begin{figure}[!ht] 
	\centering\includegraphics[width=\linewidth]{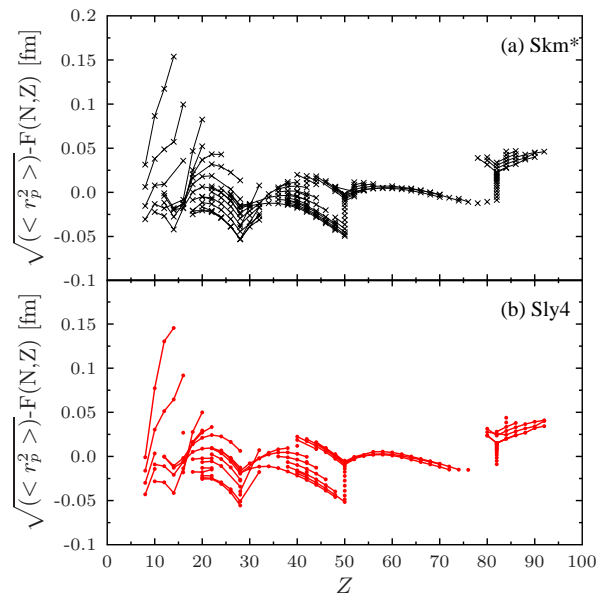}  
	\caption{(Color online) Evolution of the difference 
	 between the computed charge rms and the fitted one using formula (\ref{eq:rmsfit}). Results with the Sly4 and SkM* 
	 are respectively shown in top and bottom panel.} 
	\label{fig:comp_difrms} 
\end{figure}

We see that both functionals, even if the region of sphericity is slightly larger for SkM*, 
are globally in agreement with each other. In particular, in medium and heavy nuclei, spherical nuclei are mainly found 
around $N$ or $Z$ equal to 50 or 82 magic numbers. It should be mentioned, that in light nuclei, the spherical shape is mainly 
preferred due to pairing correlations. Indeed, when pairing is set to zero, much less nuclei are found to be spherical in their 
ground state.    

As an additional information, the root mean-square radius (rms) $\sqrt{\langle r^2 \rangle}$ of different nuclei are shown as a function of the nuclear mass $A$  in figure \ref{fig:comp_rms}. We see that both functional predicts radius that are very close from each others. This is 
consistent with the fact that their incompressibility modulus are more or less the same: $K_\infty = 216.6$ MeV (SkM*) and $K_\infty = 229.9$ 
MeV (Sly4). To further illustrate how shell effects might affect the rms, following ref. \cite{Mer94}, we have fitted the calculated 
proton rms using the 
formula:
\begin{eqnarray}
F(N,Z) & = & \sqrt{\frac{3}{5}} a_0 A^{1/3} \left( 1 + \frac{a_S}{A} - a_I \left( \frac{N-Z}{A} \right)  \right).
\label{eq:rmsfit}
\end{eqnarray}
For both functional, the fit gives $a_0 = 1.22$ fm. For Sly4 $a_S = 2.035$ fm and $a_I = 0.186$ fm while for 
SkM* $a_S = 1.9206$ fm  and $a_I = 0.189$ fm. Fig. \ref{fig:comp_difrms} displays the difference between 
the calculated charge rms and fitted value respectively for the Sly4 (bottom) and SkM* (top). This figure illustrates the impact of 
magicity and shell closure that tends to stabilize more compact shapes.

\begin{figure}[!ht] 
	\centering\includegraphics[width=\linewidth]{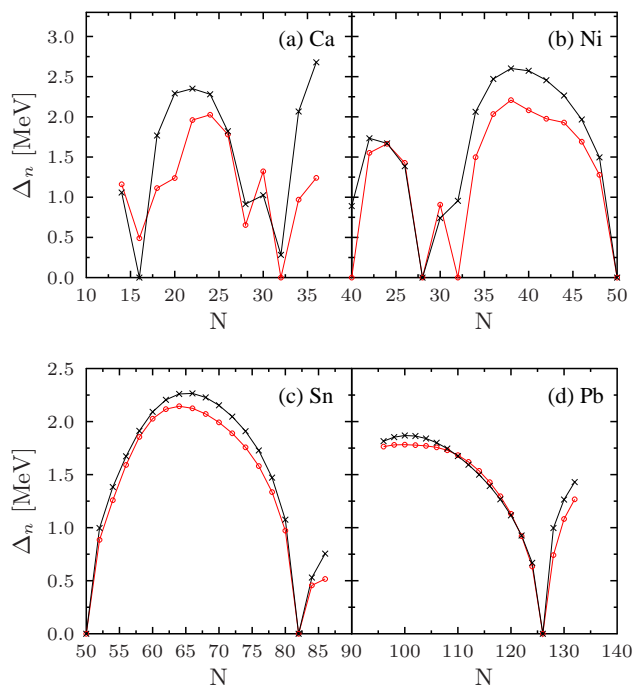}  
	\caption{(Color online) Illustration of the pairing gap in the (a) Ca, (b) Ni, (c) Sn and (d) Pb isotopes retained in the present work using the 
	 Sly4 (red circles) or SkM* (black crosses) functional. Note that the gap is shown even if the nucleus is not spherical.
	} 
	\label{fig:gap} 
\end{figure}

Finally, to characterize the pairing surface interaction, taken from ref. \cite{Ber06} and used in the present study, the neutron pairing 
gap obtained for the Ca, Ni, Sn and Pb isotopic chain is displayed as a function of the neutron number in Fig. \ref{fig:gap}.  Several comments can be made. First, these gaps are slightly higher than those reported in ref. \cite{Ter06}.  However, 
it should be noted that the BCS gap, that is a direct output from the {\rm EV8}  is different from the three (or five) points gap formula 
generally used to compare with experiments and that has been also used to adjust the pairing strength in ref.  \cite{Ber06}. 
We also see in this figure that the present surface pairing interaction leads to non-zero pairing for $^{40}$Ca.
\begin{figure*}[!ht] 
\begin{center}
	\centering\includegraphics[width=14cm ]{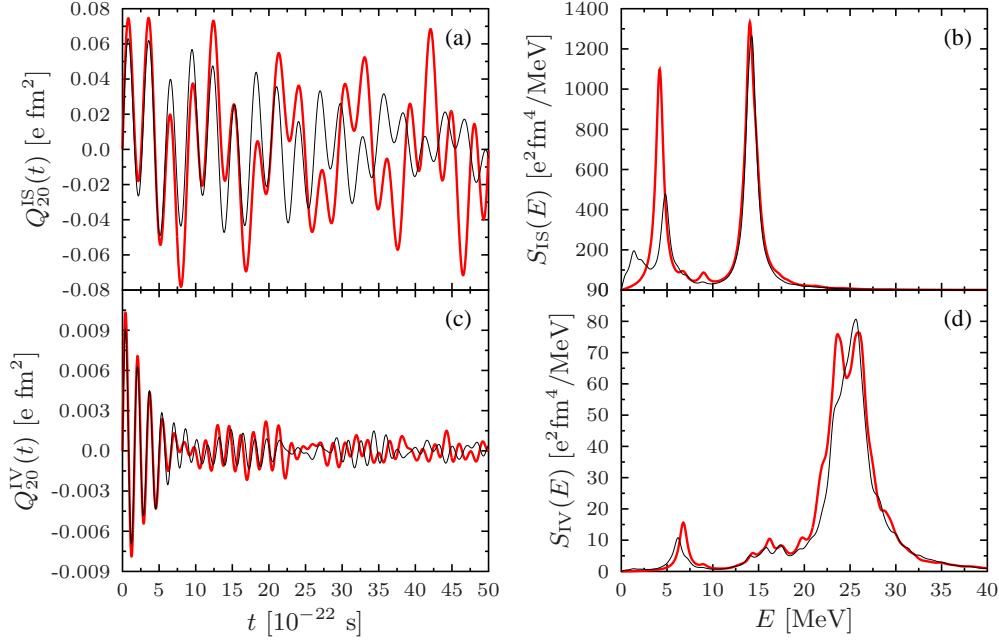}  
	\caption{(Color online) Left: Evolution of the quadrupole moments obtained using TDHF+BCS and the Sly4 functional of the isoscalar (a) and isovector (c) quadrupole moment  for a doubly closed shell nucleus $^{132}$Sn (red thick solid line) and for a neutron open shell nucleus $^{120}$Sn (thin solid line). 
	Right: corresponding response function using Eq. (\ref{eq:strength}) with a damping factor $\Gamma_0 = 1$ MeV. 
		} 
		\end{center}
	\label{fig:qt} 
\end{figure*}
Finally, we would like to mention that only nuclei that are not too exotic can be studied in the present work due to the well-known 
gaz problem appearing in the BCS framework \cite{Dob84}. This is clearly seen in the x-axis of Fig. \ref{fig:gap} that is much less extended 
than in ref. \cite{Ter06}. Nevertheless, the area of the nuclear chart explored in the present work corresponds to nuclei that can be realistically accessed in current and future nuclear reaction experiments.   

\subsection{Time-dependent evolution} 

The time evolution of each nucleus reported in figure \ref{fig:map_spher} has been systematically performed using the 3D-TDHF code \cite{Kim97}
with its BCS extension as discussed in ref. \cite{Sca13}. 
The initial condition is the spherical ground state obtained with  EV8 followed either by a quadrupole isoscalar or isovector boost.  
The time-dependent evolution is performed in r-space 
using the mesh-step parameter as in the static case and with a mesh size $L_x = L_y = 2L_z = 22.4$ fm. The numerical time-step is given by 
$\Delta t = 1.5\times10^{-24}$ s and the evolutions are followed up to a maximal time $T = 50\times10^{-22}$ s.  

As an illustration of the IS or IV quadrupole evolution, we show in Fig. \ref{fig:qt} two illustrations of nuclei: one doubly magic 
($^{132}$Sn) and one with a neutron open shell ($^{120}$Sn). The corresponding IS and IV response function are shown in right side of this figure. These evolutions illustrate the type of behavior typically observed in medium and heavy nuclei. 
From this figures, one could make the following statements:
\begin{itemize}
  \item {\it Collective energy:} Both the IS and IV response present a highly collective state at the expected energy $E \simeq 15$ MeV
  and $25$ MeV respectively.
  \item {\it Fragmentation and damping:} the IS-GQR 
displays beating between the high energy collective modes and  the ones at lower energy ($\le 5 MeV$). However the IS motion is not damped for long time. This stems from the fact that the high energy peak corresponds to a single highly collective frequency. 
On opposite,  in the IV-GQR, an effective damping is observed at short time stemming from the fragmentation of the strength around the collective energy ($\simeq 25$ MeV), that is similar to the Landau fragmentation generally observed in 
the monopole response of light nuclei. 
  \item {\it Low lying state:}  Finally, by comparing the two curves in panel (b), we see that the open shell nuclei has an extra low lying $2^+$ state  that is expected for superfluid systems (see for instance \cite{Ter06}). 
\end{itemize}

\begin{figure}[!ht] 
	\centering\includegraphics[width=\linewidth]{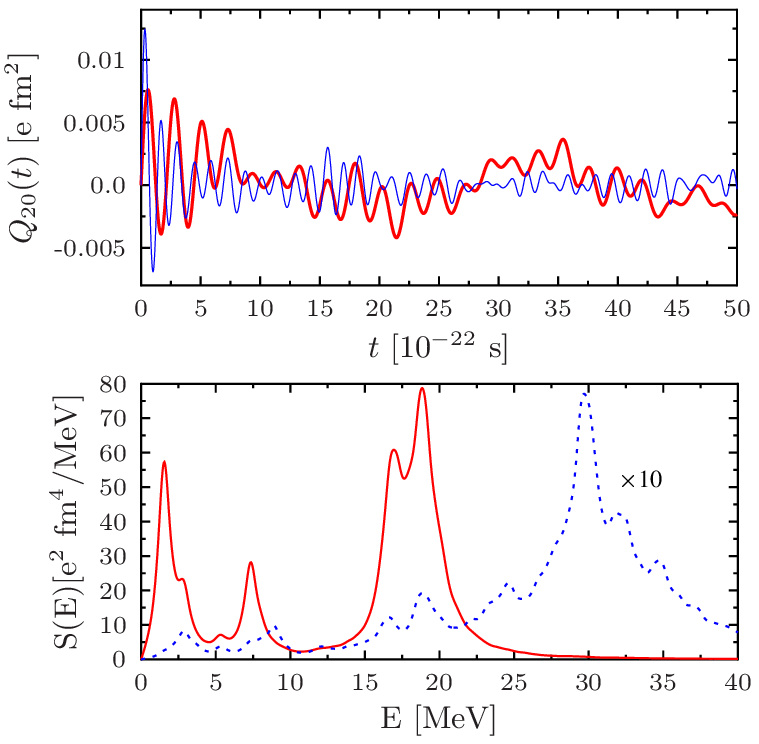}  
	\caption{(Color online) Top: Evolution of  the isoscalar (red thick line) and the isovector (blue thin line) quadrupole 
	moment   in $^{38}$S obtained using TDHF+BCS with the SkM* functional. Note that here the isovector quadrupole moment is multiplied by a factor 10.
Bottom: Corresponding strength function for the isoscalar (red thick line) and the isovector (blue dashed line) excitation.
	} 
	\label{fig:S38_IS_IV} 
\end{figure}

In Fig. \ref{fig:S38_IS_IV}, a typical evolution and response of a lighter nucleus is shown. 
For lower mass systems the fragmentation is enhanced in the IV channel while a small but non zero fragmentation 
is seen also in the IS case. We will see below that the fragmentation is seen for $A \le 100$.
The different aspects: collective energies, damping and low lying state, are discussed systematically below.

\section{Systematic study of spherical nuclei}

The different aspects discussed in previous section, have been systematically investigated. The IS- and IV- GQR strength distributions computed for all nuclei shown in Fig. \ref{fig:map_spher} can be obtained from the supplement material included with this publication 
\cite{Sup13}.

 \subsection{Energy Weighted Sum Rule}
 
 The Energy Weighted Sum-Rule (EWSR) provides a stringent test of the physical and numerical aspects. 
 These sum-rules have been extensively discussed in the literature within the Energy Density Functional approaches \cite{Boh79,Bla86} based 
 on Skyrme functional theories and we only give below the constraints that the strength should fulfill. 
 
\begin{figure}[!ht] 
	\centering\includegraphics[width=\linewidth]{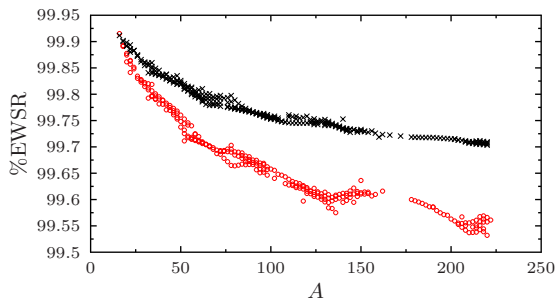}  
	\caption{(Color online) Percentage of the IS-GQR EWSR exhausted by the integral of the strength function $S_{IS}(E)$ and
obtained with 	TDHF+BCS code as a function of the mass of the nucleus. 
The reference sum-rule is computed using right side of  Eq. (\ref{eq:sum_is}). 
Results obtained with the Sly4 and with the SkM* functional are respectively shown using red circles 
and black crosses. } 
	\label{fig:sum_is} 
\end{figure}

 Starting from the specific excitation operator (\ref{eq:qis}), the EWSR for the IS-GQR is given by \cite{Har01}: 
\begin{eqnarray}
\int_0^{\infty} E S_{IS}(E) dE&=&   \frac{e^2 \hbar^2}{8\pi m} \lambda (2\lambda+1) (A - 1) \langle r^{2\lambda-2} \rangle \nonumber \\
&=& 16.501 (A-1)  \langle r^2 \rangle e^2 \rm{fm}^4 \rm{MeV}. \label{eq:sum_is}
\end{eqnarray}
In the first expression $\lambda=2$ since we are considering the GQR. The factor $(A - 1)$ instead of $A$, appear due to the 
center of mass correction  \cite{Col13}. The numerical estimates of the sum-rule (\ref{eq:sum_is}) have been made using the value $\hbar=6.58211 \times 10^{-22}$ MeV.s and $m=1.0446879 \times 10^{-44}$ MeV.s$^2$fm$^{-2}$ while the mean-radius $\langle r^2 \rangle$ are directly those computed in the static mean-field 
(figure \ref{fig:comp_rms}). The ratio between 
the integral of the strength obtained with TDHF+BCS for different nuclei 
and the right side of Eq. (\ref{eq:sum_is}) are shown in Fig. \ref{fig:sum_is}. For all nuclei considered 
here, an error lower than 0.5\%  is obtained on the total EWSR. 

\begin{figure}[!ht] 
	\centering\includegraphics[width=\linewidth]{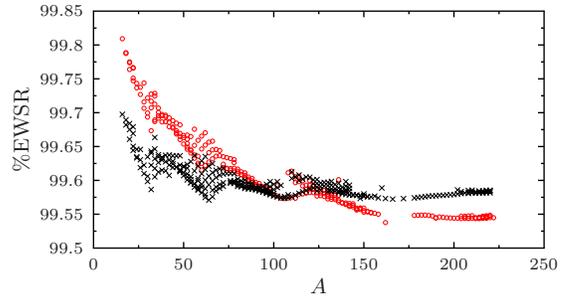}  
	\caption{(Color online) Same as Fig. \ref{fig:sum_is} for the IV-GQR.} 
	\label{fig:sum_iv} 
\end{figure}

A similar study can be made for the IV-GQR. In that case, the sum-rule is given by \cite{Sag84,Ter06,Col13}:
\begin{eqnarray}
\int_0^{\infty} E S_{IV} (E) dE&=&  \frac{e^2\hbar^2}{8\pi m} \frac{(1-A)}{A}  \lambda (2\lambda+1) \nonumber \\
&&\times \left( \frac{N^2 Z}{A^2} \langle r^{2\lambda-2} \rangle_p + \frac{N Z^2}{A^2} \langle r^{2\lambda-2} \rangle_n \right) \nonumber \\
&&+ C_{IV}(\lambda), \label{eq:sum_iv}
\end{eqnarray}
For $\lambda =2$,  $\langle r^{2\lambda-2} \rangle_{p/n} $ identifies with the proton and neutron mean radius that are again estimated directly
with EV8. $C_{IV}$ is the standard corrective term that stems from the momentum  part of the Skyrme functional. For the IV-GQR, 
we have \cite{Ter06}
\begin{eqnarray}
C_{IV}(\lambda=2) &=& \frac{e^2}{4} \left\{ t_1\left( 1+\frac{x_1}2 \right)+ t_2 \left(1+\frac{x_2}2 \right)\right\} \nonumber \\
&\times& \int \left| \nabla [ r^2 Y_{20}(\Omega)] \right|^2 \rho_n({\bf r}) \rho_p({\bf r}) d{\bf r}, 
\end{eqnarray}
where $\rho_p$ and $\rho_n$ are the neutron and proton local densities. The percentage of the EWSR exhausted by the IV 
strength and obtained with TDHF+BCS are systematically shown in Fig. \ref{fig:sum_iv}. Again, almost 100 $\%$ 
of the strength is exhausted. 

 \subsection{GQR: Mean energy and Width}
 
For each nucleus, the mean collective energy and width of the high-lying giant resonance
has been extracted by fitting the main collective peak with a Lorentzian distribution 
\begin{eqnarray}
L(E) &=&  \frac{a_0 \Gamma_C}{2\pi} \frac1{(E-E_C)^2+(\Gamma_{C}/2)^2}, \label{eq:fit}
\end{eqnarray}
where $a_0$, $E_C$ and $\Gamma_{C}$ are the fitting parameters. While $E_C$ 
is rather insensitive to the damping factor $\Gamma_0$, $\Gamma_C$ mixes physical 
effects associated to the strength fragmentation (associated with a physical width $\Gamma$) 
with the assumed smoothing parameter $\Gamma_0$.  By changing the value of $\Gamma_0$, 
it could be shown that the physical width can be simply obtained through the linear relation:
\begin{eqnarray}
\Gamma & \simeq & \Gamma_C - \Gamma_0 . \label{eq:gamma}
\end{eqnarray} 
The above 
strategy to obtain the mean energies and widths is much more precise than 
the method based on weighted moments of the strength, i.e. 
\begin{eqnarray}
E_C &\simeq& \frac{m_1}{m_0}, ~\Gamma_C \simeq   \sqrt{m_2/m_0 - (m_1/m_0)^2},
\end{eqnarray}
where $m_k$ stands for the moment of order $k$ of the strength in a restricted region of energy. We have 
found that this method is too sensitive to the integration region as well as to the smoothing parameter 
$\Gamma_0$.
 
\begin{figure}[!ht] 
	\centering\includegraphics[width=\linewidth]{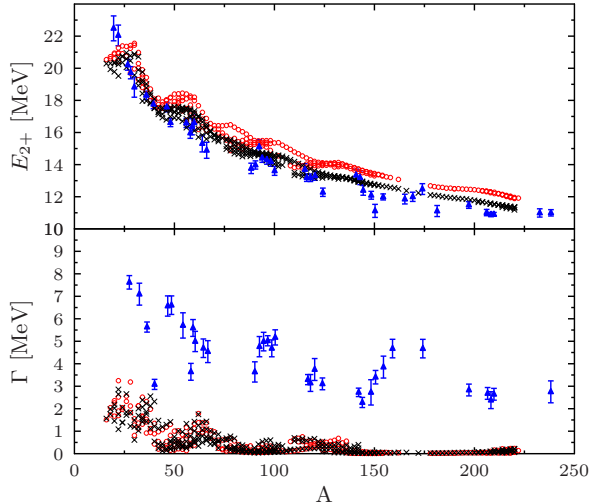}  
	\caption{(Color online) Collective energy (top) and width (middle) of the IS-GQR systematically obtained for spherical nuclei 
	using the SkM* (black crosses) and Sly4 (red circles) functionals. The blue triangles correspond to experimental datas taken from \cite{Ber81}. 
} 
	\label{fig:comp_exp_E2_IS} 
\end{figure}

The collective energy, denoted as $E_{2+}$ and the width $\Gamma$ obtained after the fit in combination with Eq. (\ref{eq:gamma}) are systematically reported as a function of the mass of the nucleus in 
 panel of Fig. \ref{fig:comp_exp_E2_IS}.  Some experimental data  taken from \cite{Ber81} are also shown as a reference.
 
Focusing first on the mean collective energy, we see that both functionals provide the correct order
magnitude of the energy over the nuclear chart with the proper $A$ dependence.  The SkM* is systematically 
closer to experimental observation. However, one should keep in mind that such a direct comparison 
should be taken with care due to the fact that 
 the technique used to get the collective energies and widths as well as the energy range considered experimentally 
 might differ from the one we used.  
 
Interestingly enough, the Sly4 functional gives collective energies that are systematically higher compared to 
the SkM* case by almost $1$ MeV  for $A \ge 70$.  It should be noted that a $1$ MeV difference, in view of desired accuracy for EDF, 
is significant enough so that the GQR might become a global criteria for the validity of a given 
functional parametrization. 
 
Bottom panel of Fig.  \ref{fig:comp_exp_E2_IS}, illustrates that the present mean-field calculation 
completely misses the fragmentation of the strength.   
The strength function is slightly fragmented for light nuclei but for medium and heavy nuclei ($A > 70$),  
the high lying collective energy essentially corresponds to a single collective energy without any spreading. 
This is clearly at variance with the experimental observation where a rather significant fragmentation is systematically observed.
This discrepancy is not surprising since a mean-field theory is expected to reproduce one-body observables but cannot really 
include two-body effects. In the case of giant resonances, the fragmentation of the strength reflects the coupling to 
complex internal degrees of freedom like the coupling to two-particles two-holes induced by in-medium collisions \cite{Gam10} or the 
coupling low lying surface modes \cite{Gam12,Bre12}. 
See for instance the extensive discussion in ref. \cite{Lac04}. As it is already known from QRPA calculation, the inclusion of pairing 
correlations does not cure this problem.

\begin{figure}[!ht] 
	\centering\includegraphics[width=\linewidth]{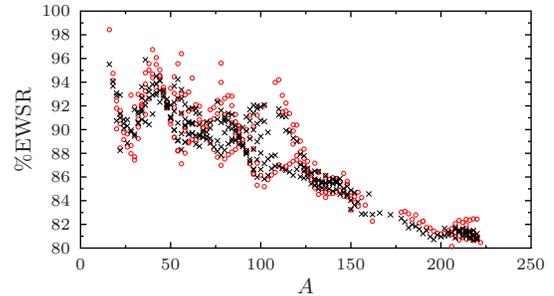}  
	\caption{(Color online) Percentage of the IS-GQR EWSR obtained by integrating the strength function 
	above 7.5 MeV shown as a function of mass for the Sly4 (red circles) and SkM* (black crosses) functionals.} 
	\label{fig:frac_ESWR} 
\end{figure}

Last, it should be noted that the collectivity of the GQR state, while always significant, continuously decreases as the mass increases.
This is illustrated in Fig. \ref{fig:frac_ESWR} where the percentage of EWSR above 7.5 MeV is shown as a function of the mass of the 
nucleus. This reduction signs the increase of the low lying components in detriment of the high lying collectivity.

\begin{figure}[!ht] 
	\centering\includegraphics[width=\linewidth]{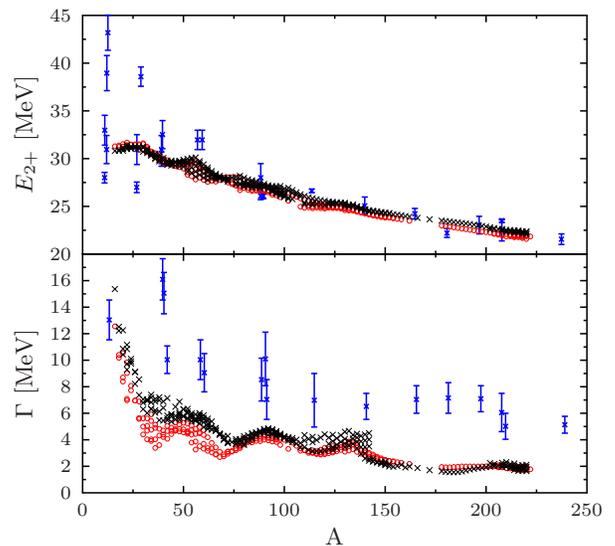}  
	\caption{(Color online) Same as figure \ref{fig:comp_exp_E2_IS} for the IV-GQR. The experimental data are taken from \cite{Sim97,Mae06}.} 
	\label{fig:comp_exp_E2_IV} 
\end{figure}
 
To complete the study, the collective energy and width of the IV-GQR is also shown in Fig. \ref{fig:comp_exp_E2_IV}. The collective energies of 
both functionals are in global agreement with the experiments except for one very light nucleus.  Regarding the width, the situation is 
slightly different compared to the IS case since the collective response is always fragmented. The fragmentation obtained with the mean-field theory accounts approximately for half of the fragmentation observed experimentally.  

It is interesting to note that the repartition of the EWSR between the low and high energy sector in the IV case
is completely opposite compared to the IS-GQR. The non negligible fraction 
of EWSR sum rule observed at low energy for light nuclei $A \le 50$ vanishes for larger mass leading to almost 100\% of the strength 
at high energy in heavy nuclei. 
  
 \begin{figure}[!ht] 
	\centering\includegraphics[width=\linewidth]{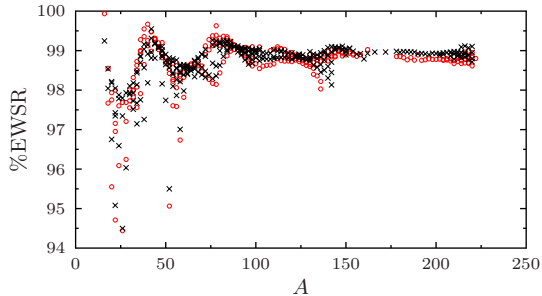}  
	\caption{(Color online) Percentage of the IV-GQR EWSR obtained by integrating the strength function 
	above 11 MeV is shown as a function of mass for the Sly4 (red circles) and SkM* (black crosses) functionals.} 
	\label{fig:frac_ESWR_IV} 
\end{figure}

\subsection{Average behavior}

The IS and IV-GQR energy can be appropriately fitted using a polynomial 
expansion in power of $A^{-1/3}$, i.e. 
\begin{eqnarray}
\overline{E_{2^+}}(A) &=& a_1 A^{-1/3} + a_2 A^{-2/3} +  a_3 A^{-1}+ ... \label{eq:pol}
\end{eqnarray}
Truncating at second order provides a rather good description of the IS-GQR but fails to 
reproduce the IV-GQR. 
In all cases, a polynomial of order 3 gives an excellent fit of the average energy 
evolution along the nuclear chart. In Table \ref{tab:fitgqr}, the set of $a_i$ parameters 
obtained for the IS and IV energies and for the two considered forces are reported. 
\begin{table}
\begin{tabular}{l |c|c|c}
\hline
 & $a_1$ (MeV)& $a_2$ (MeV)& $a_3$ (MeV)  \\
 \hline 
IS - SkM*   & 65.03 & 52.93 & -211.67 \\
IS - Sly4   &  75.50 & 3.13 & -143.69 \\
IS-exp & 72.61 & -37.45 & 0.002 \\
\hline \hline
IV - SkM*   & 177 &  -255.383 & 10.05\\
IV - Sly4   &  169.36 & -223.16 & -13.06 \\
IV -exp & 165.19 & -182.09 & -37.24 \\
\hline
\end{tabular}
  \caption{Parameters deduced for the fit of the IS and IV main collective energy with a third order polynomial in 
  $A^{-1/3}$, Eq. (\ref{eq:pol}).}\label{tab:fitgqr}
\end{table}
 \begin{figure}[!ht] 
	\centering\includegraphics[width=\linewidth]{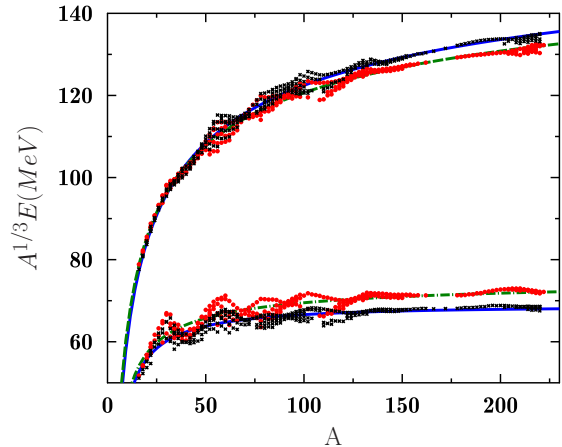}  
	\caption{(Color online) Evolution of the quantity $A^{1/3} E_{2^+}(A)$ as a function of mass. The IS (lowest energy) and IV (highest energy) energies 
	are reported for the SkM* and Sly4 functional using respectively black crosses and red filled circles. Result of the fit using 
	a third order polynomial with parameters reported in table \ref{tab:fitgqr} are shown with blue solid line and green 
	dot-dashed line for the 
	SkM* and Sly4 functional respectively  } 
	\label{fig:fitgqr} 
\end{figure}
Results of the fit are shown in Fig. \ref{fig:fitgqr} where the quantity $A^{1/3} E_{2^+}(A)$ is displayed as a function of the mass A for the two Skyrme functionals. For the sake of completeness, we also give 
in table \ref{tab:fitgqr}, the coefficients obtained by fitting the experimental points displayed 
resp. in Fig. \ref{fig:comp_exp_E2_IS}  and \ref{fig:comp_exp_E2_IV}  using expression  (\ref{eq:pol}).

\subsection{Shell effect on the GQR response}

\begin{figure*}[htbp] 
	\centering\includegraphics[width=0.8\linewidth]{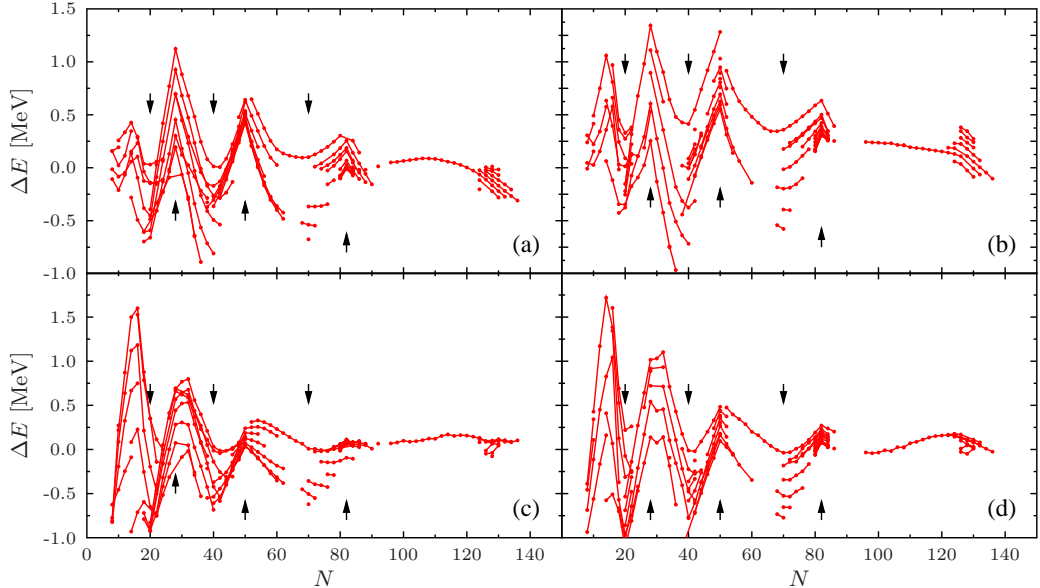}  
	\caption{(Color online) Local fluctuations $\Delta E$ of the IS-GQR (panel (c) and (d)) and the IV-GQR (panel (a) and (b)) as a function of the neutron 
	number N
	using Sly4 (panel (b) and (d)) and SkM* (panel (a) and (c)). The up-arrows indicate the magic numbers (including spin-orbit effect)
	(N=28, 50, 82) while the down-arrows correspond to the original harmonic oscillator magic numbers without the spin-orbit (N=20, 40, 70).  } 
	\label{fig:E2_sly4_shell_N} 
\end{figure*}

Shell effects are expected to induce small oscillations around 
the average dependence of the GQR energies as a function of mass. 
Thanks to the previous analysis, the local fluctuations 
of the mean energy can be de-convoluted from the average by considering the 
quantity 
\begin{eqnarray}
\Delta E(A) & = & E_{2+} (A) - \overline{E_{2^+}}(A), 
\end{eqnarray} 
where $E_{2+} (A) $ is the collective energy while $\overline{E_{2^+}}(A)$ is computed 
from formula (\ref{eq:pol}). This quantity is systematically reported as a function of the neutron number $N$ in Fig. \ref{fig:E2_sly4_shell_N} 
for the IS- and IV-GQR. Similar curves (not shown) are obtained when $\Delta E$ is plotted as a function of the proton number $Z$.

As expected,  
we do observe several maximums that exactly match
the location of shell closure at magic numbers ($N=28$, $50$ and $82$). This is due to the appearance 
of larger gaps in single-particle effective energies inducing larger energies of particle-hole excitations from which the collective states are 
built up.  Interestingly enough the deviation $\Delta E$ also presents some minimums that turn out
to be exactly located at the position of the harmonic oscillators magic numbers without spin-orbit ($N=20$, $40$ and $70$). As we will see below, this effect could be understood 
as a competition between the conserved parity during the excitation and the pairing correlations.   

\section{Pairing effect on the GQR response}

In ref. \cite{Sca13}, it was shown that one of the main effect of pairing on dynamics is to induce 
partial occupation numbers of single-particle orbitals. This effect can already be accounted 
for at the mean-field level without pairing by considering a statistical ensemble of independent 
particles instead of a Slater determinant.  In this case, considering an open shell nucleus, the occupation numbers 
of the last occupied shell are not anymore zero or one but are equal to the number of nucleons in the shell 
divided by the shell degeneracy, this is the so-called equal-filling approximation. For
the reason discussed in ref. \cite{Sca13}, this approximation can be regarded as the no-pairing reference for
TDHF+BCS.  

\begin{figure}[!ht] 
	\centering\includegraphics[width=\linewidth]{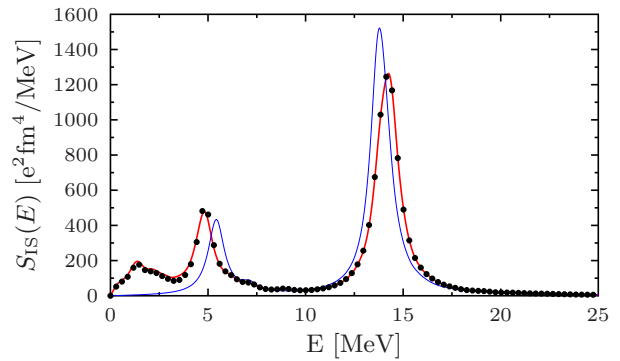}  
	\caption{(Color online) Comparison between the IS-GQR response obtained with Sly4 using the TDHF+BCS (red solid line), the TDHF+BCS with FOA (black filled circles)  and TDHF (blue thin line) for the $^{120}$Sn nucleus.	
	} 
	\label{fig:comp_FOA_TD_HF} 
\end{figure}

\begin{figure}[!ht] 
	\centering\includegraphics[width=\linewidth]{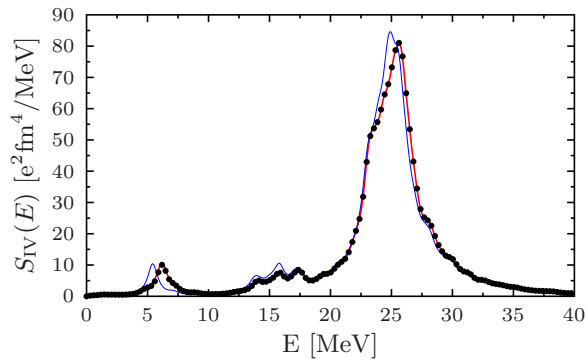}  
	\caption{(Color online)	Same as figure \ref{fig:comp_FOA_TD_HF} for the IV-GQR.
	} 
	\label{fig:comp_IV} 
\end{figure}

In Fig. \ref{fig:comp_FOA_TD_HF} and  \ref{fig:comp_IV}, the IS-GQR and IV-GQR responses are displayed using different approximations for the 
evolution. Two TDHF+BCS are considered, one where the occupation numbers and correlations are frozen (Frozen Occupation Approximation [FOA]) in time and one where the full TDHF+BCS equation of motion are solved in time. The FOA has the advantage to 
respect the continuity equation contrary to TDHF+BCS \cite{Sca12}. 
Note that since here we consider a single system, this drawback of TDHF+BCS is not crucial.  The response function with pairing 
are compared with the TDHF+equal-filling approximation case.  

The first remarkable aspect, is that the two calculations with pairing, i.e. the complete one and the FOA, are exactly on top of each others.
This confirms that in the TDHF+BCS limit, the main effect of pairing on small amplitude vibrations stems from 
the initial ground state correlations.

\begin{figure}[!ht] 
	\centering\includegraphics[width=\linewidth]{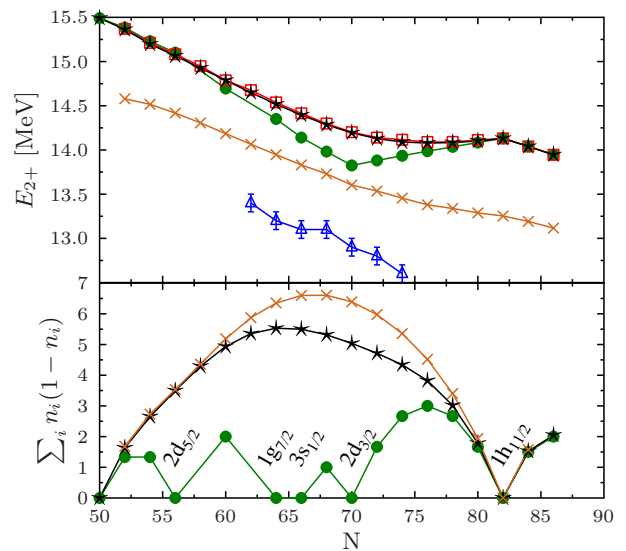}  
	\caption{(Color online) Top: Evolution of the IS-GQR energy for the Sn isotopic chain as a function of the neutron number.
Bottom: The quantity $\sum_i n_i(t_0) (1-n_i(t_0)))$  is shown. This quantity gives a measure of the fragmentation 
of single-particle occupations numbers around the Fermi energy. The TDHF+BCS (red square), the FOA (black stars) and TDHF with the equal-filling approximation (green circle) using the Sly4 functional are shown.
The TDHF+BCS results with SkM* (brown crosses) are also shown for comparison. 
	 In top panel, the experimental values from \cite{Li10} (blue open triangles) is also shown for comparison.
 } 
	\label{fig:comp_BCS_ef_Sn} 
\end{figure}

Let us now consider the difference between the TDHF+BCS case and the no-pairing limit. We focus 
first on the high energy region ($E \ge 10$ MeV). To quantify further the pairing effect on high lying states, 
the main peak collective energy of the IS-GQR is systematically reported for the Sn isotopes in top panel of 
Fig. \ref{fig:comp_BCS_ef_Sn} and for the three types of evolutions. 

Since at shell closure, $N=50$ and $N=82$, the HF-BCS ground state identifies with the HF solution, the three theories 
are exactly on top of each other. As N goes from $N=50$ to $N=82$, all theories predicts a decrease of the GQR energy consistent with the expected $A^{-1/3}$ dependence.  However, while the theory with pairing presents a smooth U shape, the TDHF approach is slightly lower   
and has a minimum at $N=70$, that corresponds to one of the magic number of the Harmonic Oscillator (HO) without spin-orbit.  This anomaly observed in the TDHF case, is located at the HO shell closure corresponding to the $(1g,2d,3s)$ shell, can be directly assigned to a parity effect. Indeed, the collective excitations have definite parities, here $\pi = +1$. 
When the number of neutrons increases, the parity of occupied single-particle states 
change from positive  (below $N=70$) to negative (above $N=70$) leading to a sudden change in the particle-hole state 
on which the collective excitation is built up.  The U shape observed in the TDHF+BCS case can be seen 
as a fossil signature of the parity effect. Indeed, due to pairing correlations, occupation numbers are more fragmented around the Fermi energy. In Fig. \ref{fig:comp_BCS_ef_Sn},  
the quantity $\sum_i n_i (1-n_i)$ is shown. This quantity measure the fragmentation of single-particle occupation numbers and falls down to zero when all occupation numbers 
are equal to zero or one. This is the case, in the BCS approximation at (spin-orbit) magic number or in the 
TDHF case when a sub-shell is fully occupied.  Again, except around magic numbers, the fragmentation is more pronounced in the BCS compared to the TDHF case. In particular, around $N=70$, single-particle states are partially occupied in BCS. Accordingly, the anomaly observed in TDHF is reduced but still occurs trough the U shape. For the SkM* functional the U shape is less pronounced because the fragmentation is larger than with Sly4.  As seen in figure \ref{fig:E2_sly4_shell_N}, the U shape that results from the competition between parity conservation and paring is observed all along the mass table (down arrows) but tends to be less pronounced as 
the mass increases.

In any case, we see that the difference between the case with or without pairing is less than the difference observed using 
different functionals (Sly4 and SkM* here). 
Our conclusion is that the main source of uncertainty in the prediction of high lying GQR collective states stems
from the functional used in the mean-field channel. 
We also see that all approximations lies above the experimental values. This might appear as a failure of the theory. However, 
one should keep in mind that correlations beyond the present approach, that account for instance for ph-phonon coupling are 
expected to shift most likely down the collective frequencies \cite{Lac04}.

\subsubsection{Systematic of low lying $2^+$ mode }

Pairing is known to play a crucial role on the low lying $2^+$ states (see for instance discussion in ref. \cite{Ber12}) at least due to the induced fragmentation of the single-particle occupation numbers near the Fermi energy. While the high energy sector of the response
are only slightly dependent on the treatment of (i) pairing effects (ii) dynamical reorganization of single-particle occupation numbers, the situation is completely different in the low energy part.  Fig \ref{fig:comp_FOA_TD_HF} illustrates that calculations whithout pairing (TDHF with equal-filling) presents a single peak at low energy around $E\simeq 6$ MeV, the full TDHF+BCS and the FOA dynamics display a more fragmented strength at low energy. In particular, new peaks are present at very low energy.

In order to have the first $2^+$ energy and the corresponding transition probability, we used a different excitation operator that acts only on protons,
\begin{eqnarray}
\hat Q=e \sum_{i=1}^{Z} r_i^2 Y_{20}(\Omega_i).
\end{eqnarray}
 \begin{figure}[!ht] 
	\centering\includegraphics[width=\linewidth]{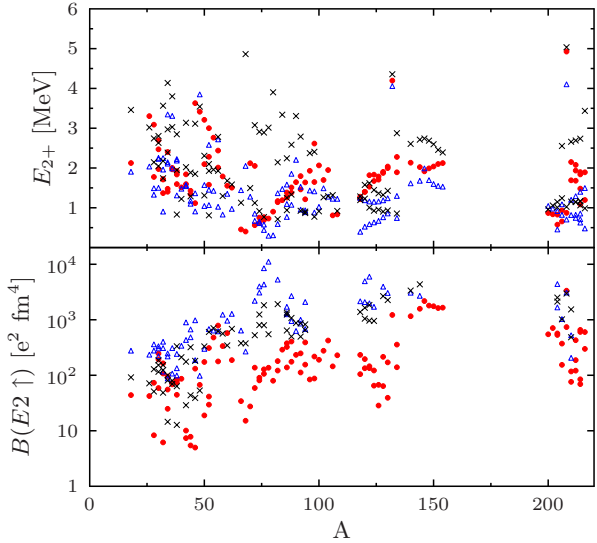}  
	\caption{(Color online) Top: Systematic comparison between the lowest $2^+$ energy obtained with TDHF+BCS using the Sly4 functional 
	(red filled circles) and the experimental first $2^+$ excited state  (blue open triangles) from \cite{Ram01}.  
       The QRPA results from \cite{Ter06,Ter08} are also shown for comparison (black crosses)} 
	\label{fig:low_ener_sly4_E2} 
\end{figure}
The energy of the lowest peak obtained with TDHF+BCS has been systematically obtained by fitting the strength using the formula \ref{eq:fit_fct} for energies $E \le 7$ MeV (see appendix \ref{app1}).
It is worth mentioning that we observed the same difficulty as in ref. \cite{Eba11} related to the convergence of low lying 
states with the model parameters. Even if the nucleus have very low $\beta_2$ values, there is a slight dependence of the response with the nucleus orientation. This dependence is reduced is the mesh parameter decreases and is not 
seen in the FOA limit. To reduce this effect, we have selected only nuclei which have $\beta_{2}<10^{-8}$. This leads to an ensemble of 112 nuclei. 

The results are reported as a function of mass in top panel of Fig. \ref{fig:low_ener_sly4_E2},
for comparison, the results of QRPA and the experimental values of the first $2^+$ are also shown. In order to quantify the agreement between the TDHF+BCS and QRPA theories, the standard deviation
\begin{eqnarray}
\sigma=\sqrt{\frac{1}{N}\sum_i^N (E^{calc}_i - E^{exp}_i)^2}
\end{eqnarray}
was computed.  For the TDHF+BCS theory  $\sigma=0.77$ MeV, while for QRPA from \cite{Ter08} $\sigma=1.00$ MeV for the same set of nuclei. This result shows that the TDHF+BCS theory can have a better predictive power for the energy of low lying states compared to QRPA even if the BCS approximation is made.

The situation is much less satisfactory for the $B(E2, \uparrow)$ values. Once the lowest peak has been fitted using Eq. (\ref{eq:fit}) 
and assuming that the physical width is zero for this peak, i.e. $\Gamma_C \simeq \Gamma_0$, then the fitted parameter $a_0$ identifies with the transition probability $|\langle\Psi_0 |\hat Q| \Psi_1 \rangle |^2$.
The 
$B(E2, \uparrow)$
value is computed through \cite{Rin80}:
\begin{eqnarray}
B(E2, \uparrow) = (2 \lambda +1) |\langle \Psi_1 | \hat Q | \Psi_0 \rangle|^2, 
\end{eqnarray}
with $\lambda=2$ for a quadrupole excitation.
The transition probability obtained in this way are systematically reported in bottom panel of Fig. \ref{fig:low_ener_sly4_E2}.
It was already noted in ref. \cite{Ter08} that the QRPA leads to a significant underestimation of the $B(E2, \uparrow)$ compared to 
experiments. The situation is even worth in the present TDHF+BCS calculation where the computed 
$B(E2, \uparrow)$ is always lower compared to the QRPA 
case sometimes by an order of magnitude. We have investigated how a change of the pairing strength, pairing interaction type (constant pairing, volume, ...) or particle number conservation through projection on good particle might affects the response.  While the high energy 
sector is rather insensitive to any of this effects, the low energy part is generally found to vary. However, at maximum a factor $2$ of increase 
or decrease is obtained that could not cure the underestimation problem. Our conclusion is that the TDHF+BCS approach is predictive for the 
energies of the low lying state but not for the $B(E2, \uparrow)$. It seems that the QRPA, and thus the TDHFB, improve the 
comparison but still miss part of the collectivity.


\section{Extraction of the density dependence of the symmetry energy from the IS- and IV-GQR}

The possibility to measure more precisely the IV-GQR in nuclei \cite{Sim97,Ich02} was recently 
proposed as a possible way to access to the symmetry energy at density lower than the saturation density \cite{Roc13}. In this reference, a precise analysis has been made on the specific $^{208}$Pb nucleus. The aim of our work is certainly not to give a detailed discussion of the symmetry energy. However, we illustrate below, how 
a systematic study made on a large scale can bring additional informations.

In the present work, we follow closely ref. \cite{Roc13}.  In this work, it has been shown that 
an approximation of the symmetry energy $S$ is given by the formula 
\begin{eqnarray}
S(A) & \simeq & \frac{\varepsilon^{\infty}_F}{3} \left\{ \frac{A^{2 /3}}{8 (\varepsilon^{\infty}_F)^2}\left[\left(E^{IV}_{2^+}\right)^2  - 2 \left( E^{IS}_{2^+} \right)^2 \right]+1 \right\}, \label{eq:isiv}
\end{eqnarray}
where $\varepsilon^{\infty}_F$ is the infinite nuclear matter Fermi energy at saturation, that is assumed here equal to $37$ MeV.  $E^{IS}_{2^+}$ and $E^{IV}_{2^+} $ are respectively the isoscalar  and isovector collective energies.
 
\begin{figure}[!ht] 
	\centering\includegraphics[width=\linewidth]{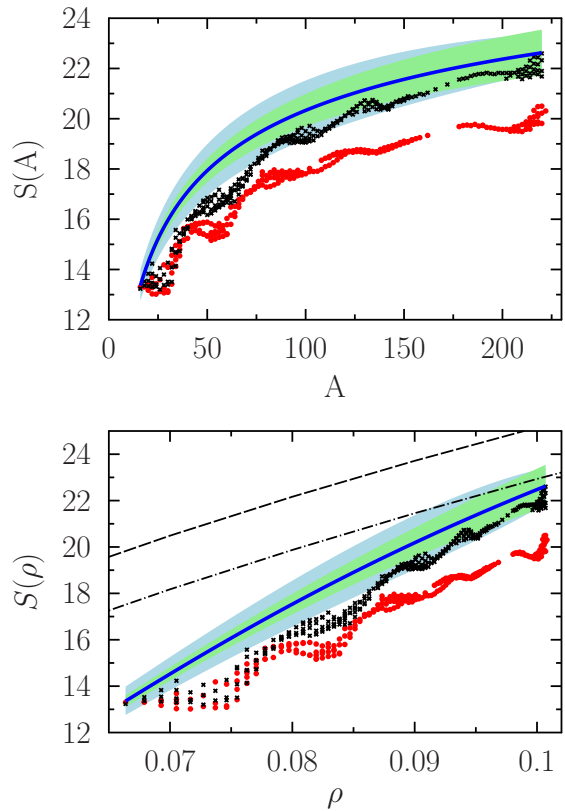}  
	\caption{(color online) Top: Symmetry energy as a function of the mass $A$ extracted using Eq. (\ref{eq:isiv}) for the Sly4 (red filled circles) and the SkM* (black crosses) cases. The blue solid line corresponds to the curve obtained 
	using the experimental data (see text for detail), the light blue area corresponds to the uncertainty
when accounting for the error bars on the IV-GQR energy solely, while the light green area 
is the uncertainty when an error bars reduced to 500 
keV is assumed.
Bottom: Same as in the top panel except that the symmetry energy is displayed as a function of 
the density. The  symmetry energy in infinite nuclear matter (Eq. (4) of Ref. \cite{Che05}) is also shown for the Sly4 (dashed line) and SkM* (dot-dashed line).} 
\label{fig:asym} 
\end{figure}

The quantity $S(A)$ is displayed in Fig. \ref{fig:asym} using the IS- and IV-GQR  main peak energy reported in 
top panel of Fig. \ref{fig:comp_exp_E2_IS} and \ref{fig:comp_exp_E2_IV}. Following ref. \cite{cen09}, the mass 
dependence can be transformed into an effective density dependence using 
\begin{eqnarray}
\rho &=& \rho_0 -\rho_0/(1 + cA^{1/3}), \label{eq:rhoa}
\end{eqnarray} 
where $c$ is set to impose $\rho=0.1$ fm$^{-3}$ for the $^{208}$Pb. Here $\rho_0$ 
stands for the saturation density that corresponds to $\rho_0 = 0.16$ fm$^{-3}$ for both functionals \cite{Bar82,Cha98}, leading to $c= 0.28$.
In bottom part of Fig. \ref{fig:asym}, the quantity $S$ obtained with Eq. (\ref{eq:isiv}) is shown as a function of the density $\rho$ deduced from the formula (\ref{eq:rhoa}).
	
Although the IV-GQR data are rather poor, similar curve can be obtained using the experimental points 
reported in Fig. \ref{fig:comp_exp_E2_IS} and Fig. \ref{fig:comp_exp_E2_IV}. To smooth out the uncertainty 
on experimental collective energies and be able to get information on a wider range of nuclei, we have used 
the polynomial fit of the experimental points, Eq. (\ref{eq:pol}),  instead of the experimental  points themselves.  
Using, parameters reported in table \ref{tab:fitgqr}, a mean tendency of the IS- and IV-GQR can be computed for all 
masses, that can be used in formula (\ref{eq:isiv}). The results is shown by solid line in Fig. \ref{fig:asym}.   
The error bars, especially on the IV-GQR are rather large (1-2 MeV). To see the effect of the IV-GQR, we also 
show in figure  \ref{fig:asym} with the light blue area the uncertainty due to such error bars. The lower and upper boundaries 
of the area are obtained by using  the IV collective energy plus or minus the energy error bars (where 
the experimental error bars have been fitted by a simple linear function). Finally, we also show on the figure 
the density dependence of the symmetry energy obtained with the Skyrme functional in infinite nuclear matter, called here after $S_{\infty}$, whose expression can be found for instance in Refs. \cite{Sto03,Che05}. 

With all these curves at hand some interesting conclusions can be drawn.
First, comparing the results obtained with the two Skyrme functionals, we see that for low masses, $A \le 70$, both 
functionals agree with each others. However, as the mass increases, a significant difference can be observed 
  in the symmetry energy extracted with (\ref{eq:isiv}). This difference can obviously be traced back in the difference 
  observed in the collective peak energies. It should be noted, that since the symmetry energy 
  is calculated as a difference of energies of collective peaks. Such differences might still lead to the same value for $S(\rho)$.
  However, we see that the IS peak in SkM* is systematically lower compared to Sly4 while for the IV peak 
  it is the opposite. Therefore, the differences in energies add them up and 
  cooperate to further show up in the finite system symmetry energy.  

  The symmetry energy extracted from the GQR are systematically lower 
  than the one obtained analytically in infinite nuclear matter. 
  In particular, analytical expressions lead to higher symmetry energies in Sly4 (dashed line) compared to SkM* (dot-dashed line) while the opposite is seen from the GQR based 
  symmetry energy. 
  It could also be noted that a rather significant difference is observed between $S_\infty (\rho)$ and the GQR based symmetry energy. Besides the absolute value, the slope is also different between the infinite system and finite system cases.  These differences clearly point out subtle finite size effects when Eq. (\ref{eq:isiv}) 
  is used, deserving more studies that are out of the 
 scope of the present work.
 
  
  Applying formula (\ref{eq:isiv}) with experimental collective energies already gives gross properties 
  of the symmetry energy as a function of mass. The main source of uncertainty is coming from the very few 
  measurements made for the IV-GQR and from their large error bars.  Nevertheless we see that the 
  experimentally deduced symmetry energy is globally in agreement with the SkM* prediction for density above 
  $0.085$ that corresponds to $A\ge 70$ while it is not with the Sly4 case. To illustrate the impact of improving 
  the measurement precision, we have artificially reduced the error bars down to 500 keV (light green filled area).    

\section{Conclusion}

The IS and IV response in spherical nuclei has been obtained over a large set of nuclei 
using the TDHF+BCS approach based on different Skyrme energy density functional.  The 
present study, is first a proof of principle that such an approach is fast and simple 
enough to make possible qualitative and quantitative systematics over the nuclear chart.
It is shown, that the TDEDF method globally reproduces the average evolution of the main collective 
energy but, due to the missing two-body effects, cannot reproduce its spreading width. 
A careful analysis, has shown that the mean-collective energy does depends rather significantly 
on the Skyrme functional. The dependence is large enough to eventually use the GQR as a criteria of
selection of the functional itself. In the present work, it is shown that the SkM* reproduces much better 
the experimental IS-GQR than the Sly4 functional. 

Besides the global effects, the local fluctuations due to shell structure effects, the parity conservation as well pairing are discussed. It is shown
that all these effects might contributes locally to the fluctuations around the average properties.  While around magic numbers, the shell stabilization induces an increase of the GQR energy, in between magic numbers, the GQR presents a U-shape that is the result of the competition between pairing and parity effects.  

The possibility to describe low lying $2^+$ state with our time-dependent approach is discussed. It is shown that the TDHF+BCS method is able to globally describe the energy of the first $2^+$ state but strongly underestimate the transition operator.   

Finally, the possible extraction of the density dependence of the symmetry energy from 
IS- and IV-GQR is discussed. We show that large scale calculation like the one presented here 
can provide interesting information on the symmetry energy but requires a better understanding 
of finite size effects and would benefit from high precision measurement of the IV-GQR.  
 
All response functions used in the present work 
can be downloaded with the supplement material \cite{Sup13}.


\appendix

\section{Strength function obtained in real time calculations}
   \label{app1}
In order to determine the peak with the lowest energy, we have developed a fitting procedure that take into account the finite time of the evolution.
Starting from the evolution of $Q(t)$,
\begin{eqnarray}
Q(t) \simeq Q(0)+i\lambda \sum_\nu |\langle \Psi_\nu |\hat Q | \Psi_0 \rangle|^2 
\left( e^{-i\frac{{E}_\nu t}{\hbar}} - e^{i\frac{{E}_\nu t}{\hbar}} \right),
\end{eqnarray}
The  Fourier transform, obtained with an integration between the time $t_0$ and the final time $T$, with a damping factor $\Gamma_C$, becomes
\begin{widetext}
\begin{eqnarray}
S_{Q}(\omega)&=& - \frac{1}{\pi \lambda \hbar} {\rm Im} \int_0^{T}  e^{-\frac{\Gamma_C}{2\hbar} t } (Q(t)-Q(0)) \sin(\omega t) dt  \nonumber \\
&=& \frac{1}{\pi} \sum_\nu^N |\langle \Psi_\nu | \hat Q | \Psi_0 \rangle|^2 
	\left\{ 
 		\frac{ - \frac{\Gamma_C}2 \left( e^{-\frac{\Gamma_C T}{2 \hbar}} \cos\left(\frac{(E-{E}_\nu)T}{\hbar}\right) -1 \right) + (E-{E}_\nu) e^{-\frac{\Gamma_C T}{2\hbar}} \sin\left(\frac{(E-{E}_\nu)T}{\hbar}\right) }
			{(\frac{\Gamma_C}{2 })^2+(E-{E}_\nu)^2} \right. \nonumber \\
&&	\quad \quad \quad \quad \quad \quad \quad 	\quad \left. - \frac{ - \frac{\Gamma_C}2 \left( e^{-\frac{\Gamma_C T}{2 \hbar}} \cos\left(\frac{(E+{E}_\nu)T}{2 \hbar}\right) -1 \right) + (E+{E}_\nu) e^{-\frac{\Gamma_C T}{2\hbar}} \sin\left(\frac{(E+{E}_\nu)T}{\hbar}\right) }
			{(\frac{\Gamma_C}{2})^2+(E+{E}_\nu)^2}   
	\right\}. \label{eq:fit_fct}
\end{eqnarray}
\end{widetext}
In the present articles, fits are performed using the above formula that accounts for both the smoothing parameter 
and the finite time interval for the evolution. 
In practice, for the low lying state, due to the possible appearance of several peaks, the fit
is made assuming up to  $N=7$ peaks in the energy region $E \le 7$ MeV.

\end{document}